\documentclass{iau} 
\usepackage{graphicx}
\usepackage{natbib}

\newcommand\aap{{A\&A}}%
\newcommand\araa{{ARA\&A}}%
\newcommand\mnras{{MNRAS}}%
\newcommand\apj{{ApJ}}%
\newcommand\apjl{{ApJ}}%
\newcommand\aj{{AJ}}%
\newcommand\nat{{\it Nature}}%
\title[IL Spectroscopy of Extragalactic GCs] %% give here short title %%
{Integrated Spectroscopy of Extragalactic Globular Clusters}

\author[Charli M. Sakari]   %% give here short author list %%
       {Charli M. Sakari$^1$
         \thanks{Present address: Department of Physics \& Astronomy, San Francisco State University, 1600 Holloway Avenue, San Francisco, CA 94132.}}

\affiliation{$^1$Department of Astronomy, University of Washington, Seattle WA
98195-1580, USA \\ email: {\tt sakaricm@uw.edu}}

\pubyear{2019}
\volume{351}  
\jname{Star Clusters: From the Milky Way to the Early Universe}
\editors{A. Bragaglia, M.B. Davies, A. Sills \& E. Vesperini, eds.}
\begin{document}

\maketitle
%. CONTINUE EDITING FROM HERE

\begin{abstract}
Integrated light (IL) spectroscopy enables studies of stellar
populations beyond the Milky Way and its nearest satellites. In this
paper, I will review how IL spectroscopy reveals essential information
about globular clusters and the assembly histories of their host
galaxies, concentrating particularly on the metallicities and
detailed chemical abundances of the GCs in M31.  I will also briefly
mention the effects of multiple populations on IL spectra, and how
observations of distant globular clusters help constrain the source(s)
of light-element abundance variations.  I will end with future
perspectives, emphasizing how IL spectroscopy can bridge the gap
between Galactic and extragalactic astronomy.
%% add here a maximum of 10 keywords, to be taken form the file <Keywords.txt>
\end{abstract}

\firstsection % if your document starts with a section,
              % remove some space above using this command.
\section{Introduction}
The chemical abundances, ages, and velocities of globular clusters
(GCs) provide essential information about their host galaxies,
particularly for distant galaxies whose individual stars are too faint
to observe.  A GC's metallicity and abundance ratios (including
[$\alpha$/Fe]) are set by the mass and star formation history of its
original host galaxy.  Massive galaxies, for example, can produce more
metal-rich stars and GCs than lower-mass galaxies.  Similarly, the
[$\alpha$/Fe] ratios (represented here by [Ca/Fe]) indicate the
relative contributions from massive and low-mass stars, specifically
the contributions from Type II core-collapse supernovae relative to
the Type Ia supernovae that are due to the detonation of a white
dwarf. The combination of [Fe/H] and [Ca/Fe] can be a powerful
diagnostic for revealing the mass of a GC's host galaxy (e.g.,
\citealt{Tolstoy2009}).  In a large galaxy with hundreds of GCs, this
information provides insight into the low-mass systems that have built
up the larger galaxy's halo over cosmic time.

For distant GCs, abundances, ages, and velocities can all be obtained
from integrated light (IL) spectroscopy, where a single spectrum is
obtained for an entire cluster.  IL spectroscopy is necessarily more
complicated that analyses of individual stars, as it requires some
assumptions about the underlying stellar populations.  However, IL
spectroscopic techniques have been tested extensively and validated
with IL observations of Milky Way GCs (e.g.,
\citealt{Schiavon2002a,Schiavon2002b,Schiavon2004,MB08,Sakari2013,Sakari2014,Colucci2017,Larsen2017}).
The spectral resolution dictates the quantities that can be derived
from an IL spectrum.  At low-resolution spectral lines are blended
together, making it harder to determine individual abundances for most
elements; at high-resolution, individual lines can be resolved,
yielding high-precision abundances of individual elements.
Lower-resolution spectra can provide ages, metallicities, velocities,
and some abundances, depending on wavelength coverage, the metallicity
of the target, and the quality of the spectra; lower resolution IL
spectroscopy requires less observing time than high-resolution
spectroscopy, and can therefore be applied to many
targets. Alternatively, high-resolution spectra provide detailed
abundances of a wide variety of individual elements, particularly from
weaker lines that can be very difficult to obtain at lower resolution
(e.g., Na or Eu) or that become prohibitively weak at low metallicity,
but only for the brightest targets.  The choice of spectral resolution
is therefore motivated by the science goals.

IL spectroscopy at various resolutions has been applied to many
extragalactic systems over the last few decades.  This paper focuses
on the GCs in M31, the Andromeda Galaxy.  As the closest large spiral
galaxy to the Milky Way, M31 is an ideal place to begin discussions of
IL spectroscopy.  Its proximity means that its brightest individual
stars can be resolved, though they are still generally too faint for
spectroscopy.  These observations of GCs in the inner and outer
regions will provide insight into M31's assembly history; these
observations also probe GCs that are unlike those in the Milky Way,
providing insight into the formation of multiple populations in GCs.

\section{The Globular Cluster System of M31}
\begin{figure}[b]
% \vspace*{-2.0 cm}
  \begin{center}
    \centering
 \includegraphics[width=5.75in]{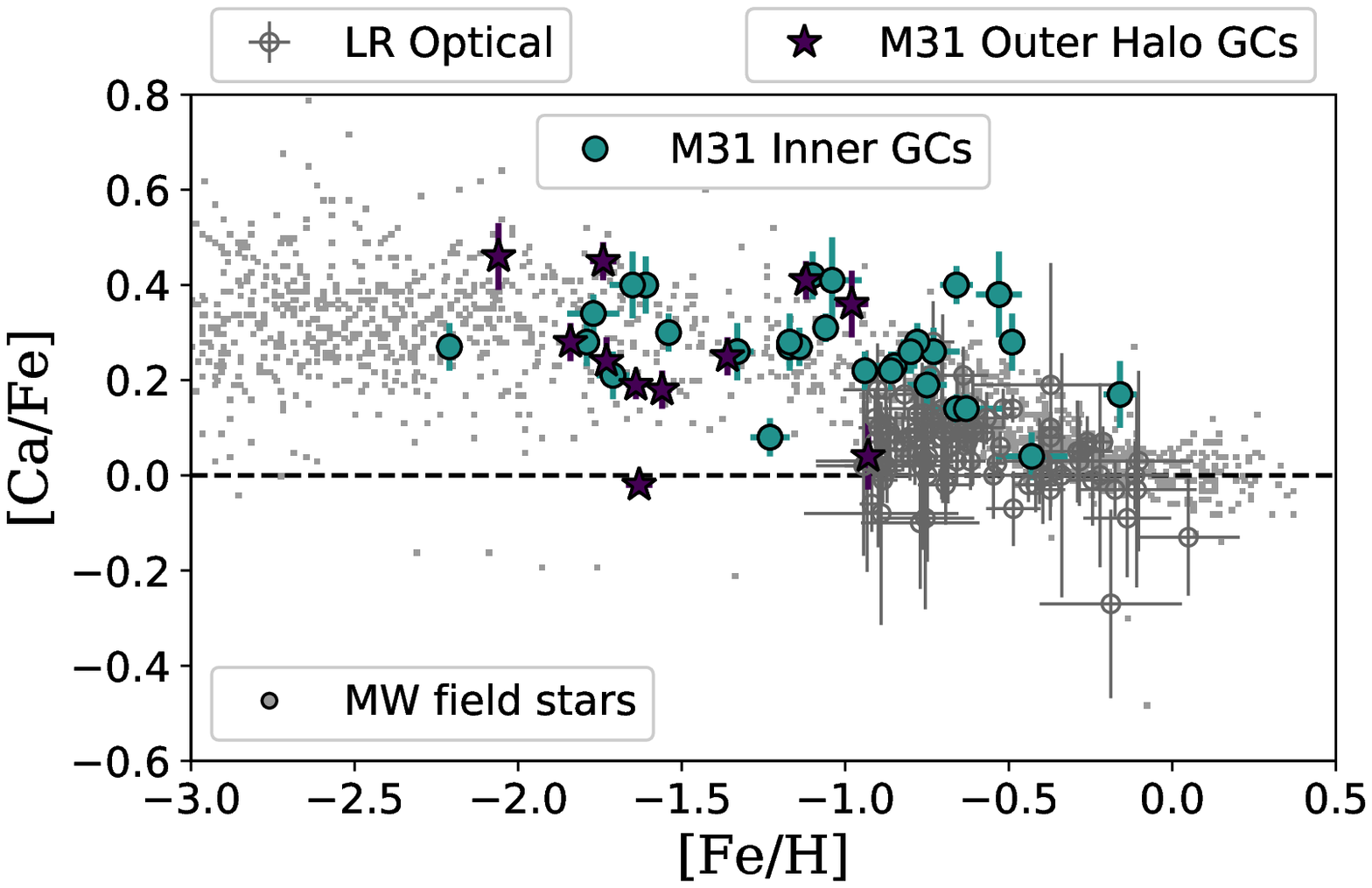} 
% \vspace*{-1.0 cm}
 \caption{[Ca/Fe] ratios as a function of [Fe/H] in
   Milky Way field stars (grey points; from
   \citealt{Venn2004,Reddy2006,Sakari2018}) and M31 GCs.  The open
   grey circles show low-resolution IL measurements from
   \citet{Schiavon2013}; the sample is limited to metal-rich clusters
   ($[\rm{Fe/H}]>-0.95$) because of difficulties modeling the
   horizontal branch in lower metallicity clusters.  The cyan circles and
   purple stars show high-resolution IL ratios for clusters in the
   inner ($R_{\rm{proj}}<25$ kpc) and outer halo, respectively, from
   \citet{Colucci2014} and \citet{Sakari2015,Sakari2016}.}
   \label{fig:CaFe}
\end{center}
\end{figure}

There have been many spectroscopic observations of M31 GCs, dating all
the way back to 1932, when \citet{Hubble} reported the radial velocity
of a single cluster.  Since then, several groups have conducted
homogeneous analyses of large sample of GCs or detailed studies of
individual GCs.  Many GCs have been studied in the inner regions (with
projected distances from the center of M31 of $R_{\rm{proj}}<25$ kpc),
where GCs are found in the disk, the bulge, and the inner halo.
Studies of large samples of GCs have found that M31 contains many more
GCs than the Milky Way, including a population of more massive GCs
\citep{Strader2011}, a population of younger GCs
\citep{Caldwell2009,Caldwell2011}, an excess of metal-rich GCs
\citep{Caldwell2011}, and a significant fraction of GCs that rotate
with the galaxy \citep{Caldwell2016}.  The presence of
so many GCs, especially so many that are unlike Milky Way GCs,
hints that M31's star formation and accretion history has been more
intense than the Milky Way's.  However, the chemical abundances of
many of these inner GCs, particularly the [Ca/Fe] ratios
(\citealt{Schiavon2013,Colucci2009,Colucci2014,Sakari2016}; see Figure
\ref{fig:CaFe}) reveal a GC population that is chemically similar to
the Milky Way.  The vast majority of the inner GCs are consistent with
{\it in situ} formation in M31 or accretion from a fairly massive
dwarf galaxy.

Images of M31's outer halo ($R_{\rm{proj}}>25$ kpc) from the
Pan-Andromeda Archaeological Survey (PAndAS;
\citealt{McConnachie2009}) reveal a number of stellar streams, which
are debris from dwarf satellite galaxies that are being accreted into
M31. There are 92 GCs known in the outer halo  of M31
\citep{Mackey2019}, many of which were discovered in PAndAS and
proceeding surveys.  On average, the outer halo GC population is
generally more metal-poor than the innermost GCs, as expected from the
overall metallicity gradient in the field stars
\citep{Gilbert2014,Ibata2014}. However, there are several fairly
metal-rich GCs at large projected distances from the center
\citep{Colucci2014,Sakari2015,Wang2019}, which may have been
accreted. \citet{Mackey2019} also found that $35-60$\% of the outer
halo GCs are likely to be associated with bright stellar streams,
based on their locations and kinematics (also see
\citealt{Veljanoski2014}). These streams and GCs were therefore likely
accreted within the last few Gyr.  The [Fe/H] and [Ca/Fe] ratios of
several of these GCs place constraints on the masses of the dwarf
galaxies that have created these streams (see \citealt{Sakari2015}).
For example, the high [Ca/Fe] abundance of the GC H10, at
$[\rm{Fe/H}]\sim -1.4$ \citep{Sakari2015}, suggests that it formed in
a fairly massive dwarf galaxy; H10 might also be a member of the SW
Cloud, a bright feature to the southwest of M31 \citep{Mackey2019}.
The chemical abundances of H10 therefore suggest that the progenitor
galaxy of the SW Cloud was more massive than the current stream
indicates, in agreement with other papers (e.g.,
\citealt{McMonigal2016}).

Several GCs that are not associated with any substructure look to have
similar chemical abundances as dwarf galaxy stars. PA-17 has
$[\rm{Fe/H}]\sim-0.9$ and $[\rm{Ca/Fe}]\sim0$, a signature of
metal-rich stars in the a galaxy with a similar mass as the Large
Magellanic Cloud \citep{Sakari2015}.  Given its high metallicity, it
is possible that PA-17 originated in the Giant Stellar Stream (GSS),
the feature that contains the majority of the metal-rich stars in the
outer halo \citep{Ibata2014}.  Although PA-17 does not currently lie
on the GSS, models suggest that the GSS has multiple wraps around the
galaxy (e.g., \citealt{Fardal2013})---in this case, PA-17 could have
been stripped from the GSS during an earlier encounter.
\cite{Colucci2014} also found the more metal-poor GC, G002, to have
$[\rm{Fe/H}]\sim -1.6$ and $[\rm{Ca/Fe}]\sim0$, similar to the
metal-rich GC in the Fornax dwarf spheroidal galaxy
\citep{Hendricks2016}. The lack of substructure surrounding these two
GCs indicates that they were unlikely to have been accreted very
recently.

Ultimately, radial velocities, ages, metallicities, and detailed
abundances of individual elements can all be derived from IL spectra
of extragalactic GCs, providing valuable constraints on the birth
sites of the GCs.  Future observations will continue to shed light
on M31's assembly history.

\section{Multiple Populations in M31 GCs}
\begin{figure}[b]
% \vspace*{-2.0 cm}
  \begin{center}
    \centering
 \includegraphics[width=5.75in]{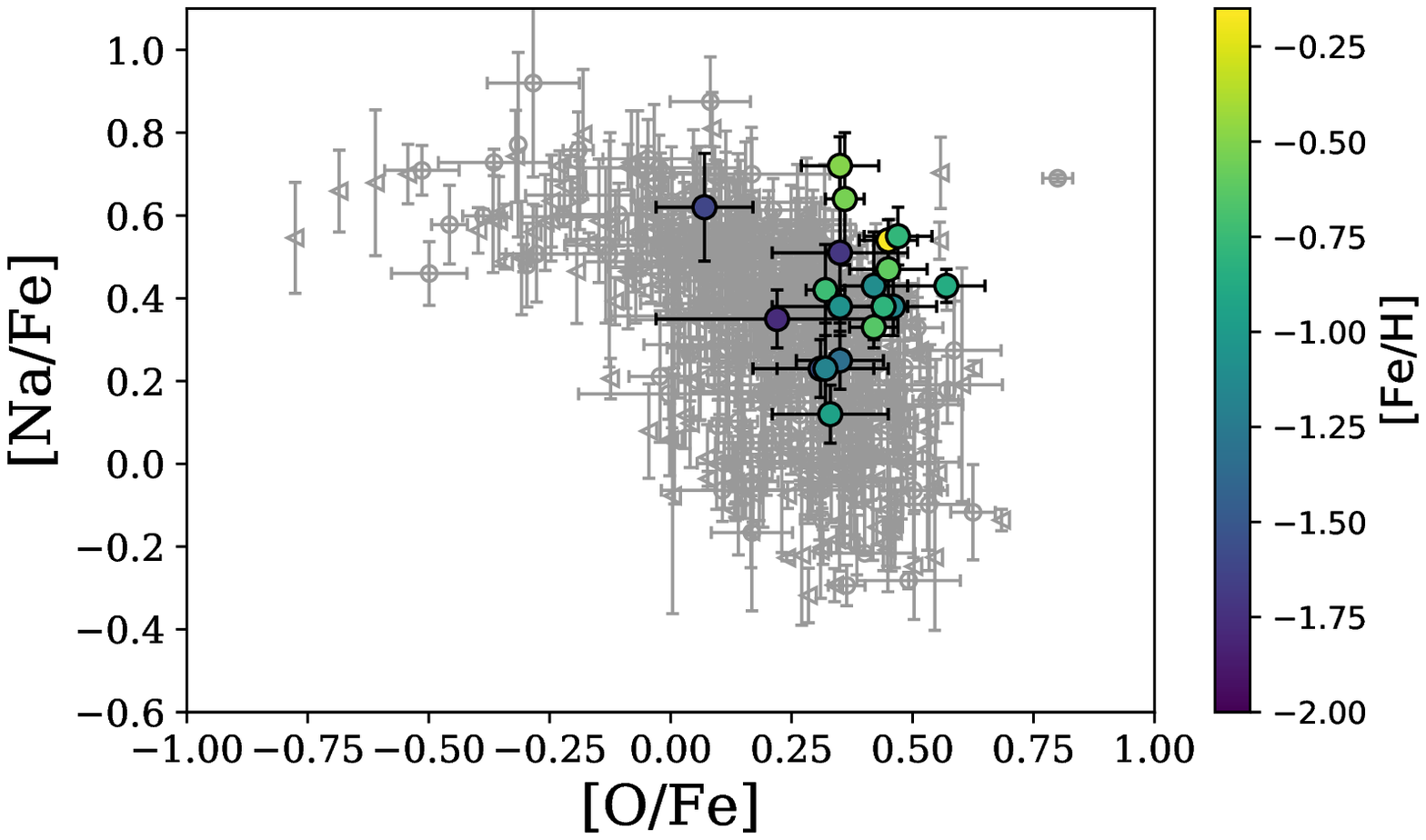} 
% \vspace*{-1.0 cm}
 \caption{IL [Na/Fe] vs. [O/Fe] ratios for a sample
   of M31 GCs, color-coded by [Fe/H].  The [Na/Fe] ratios come from the
high-resolution, optical observations of \citet{Colucci2014} and
\citet{Sakari2016}, while the [O/Fe] ratios are from the analysis of
$H$-band spectra by \citet{Sakari2016}.  The grey open circles show
individual stars from \citet{Carretta2009}, while the sideways
triangles show upper limits in [O/Fe].}
   \label{fig:NaO}
\end{center}
\end{figure}

Light element abundance variations, commonly referred to as ``multiple
populations,'' are a ubiquitous feature in all classical GCs in the
Milky Way and its nearby satellites (see \citealt{BastianLardo2018}).
IL spectroscopy provides the opportunity to assess the presence of
multiple populations in distant GCs.  Since IL spectra provide only a
single abundance for an entire GC, star-to-star spreads cannot be
directly detected, though they can be inferred.  Several papers have
demonstrated that M31 GCs show high IL [N/Fe], [Na/Fe], or [Al/Fe],
indicating the presence of multiple populations
\citep{Colucci2009,Colucci2014,Schiavon2013,Sakari2015,Sakari2016,Larsen2018}.
Figure \ref{fig:NaO} shows IL [Na/Fe] vs. [O/Fe] ratios for a sample
of M31 GCs, color-coded by [Fe/H].  Many of these GCs fall into the
``intermediate'' region defined by \citet{Carretta2009}, which falls
between the ``primordial'' region (where stars have similar Na and O
as field stars) and the ``extreme'' region (where stars are highly
enhanced in Na and deficient in O, characteristic of hot H-burning).
The location of the M31 GCs in this plot suggests that they host a
significant population of Na-enhanced stars.

\begin{figure}[b]
% \vspace*{-2.0 cm}
  \begin{center}
    \centering
 \includegraphics[width=5.75in]{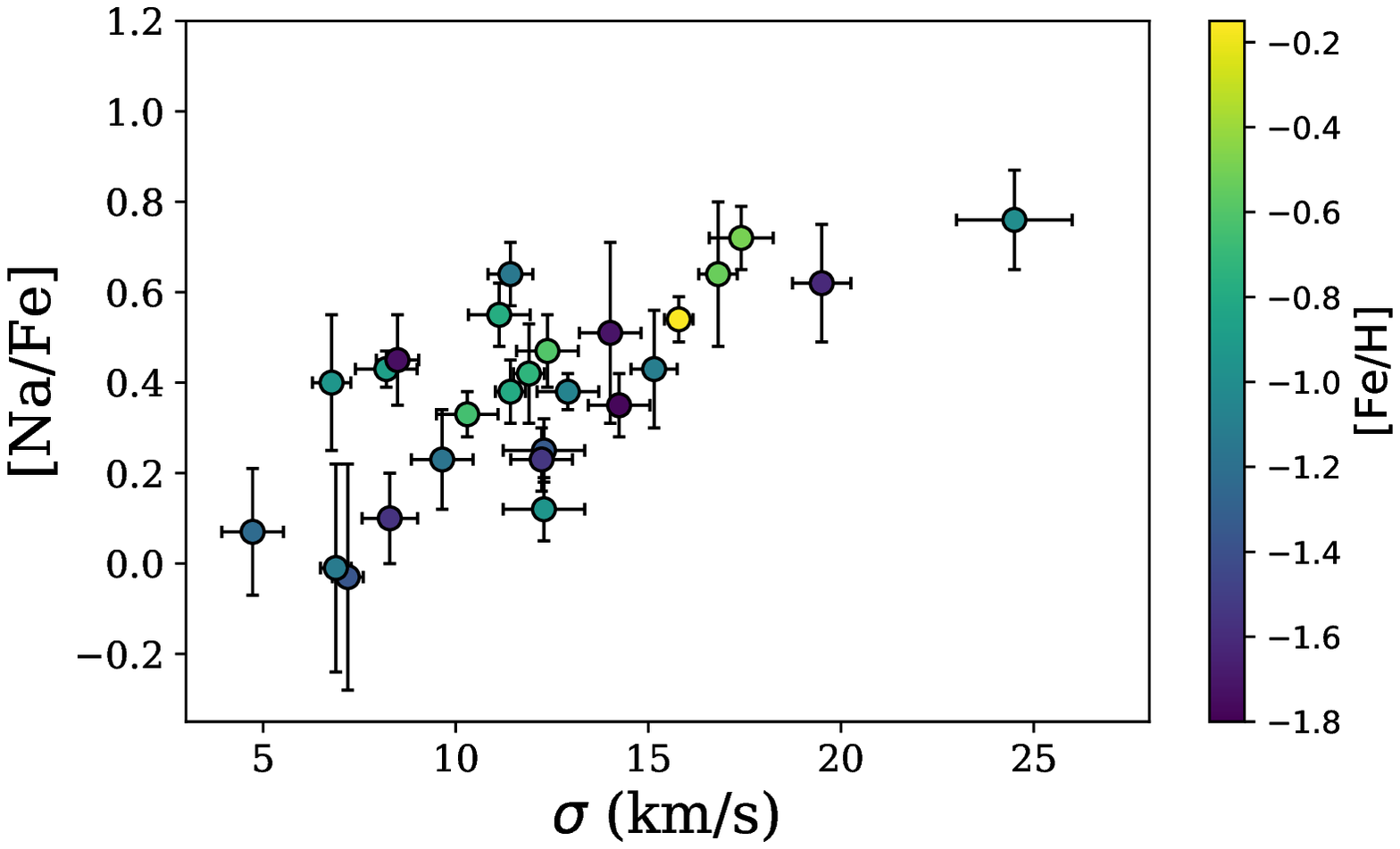} 
% \vspace*{-1.0 cm}
 \caption{IL [Na/Fe] (from \citealt{Colucci2014},
   \citealt{Sakari2016}, and Sakari et al. {\it in prep.}) versus
   cluster velocity dispersion.  The points are color-coded by [Fe/H].}
   \label{fig:NaMass}
\end{center}
\end{figure}

Clusters with a greater proportion of ``intermediate'' or ``extreme''
stars should have higher IL [Na/Fe] ratios.  The IL abundances
therefore provide an opportunity to investigate trends with cluster
properties, such as mass.  \cite{Schiavon2013} found a significant
increasing trend in the IL [N/Fe] ratios with cluster mass.  A similar
trend is also evident in IL [Na/Fe] with cluster velocity dispersion
(Figure \ref{fig:NaMass}), one that extends to the massive cluster G1
(Sakari et al. {\it in prep.}).  Similar trends in [Na/Fe] were also
found by \citet{Colucci2014} and \cite{Sakari2016}; note that both
papers also found hints of similar trends in Mg and Al, but only for
some clusters.  Altogether, the IL abundance ratios suggest that more
massive clusters are able to produce relatively more ``intermediate''
or ``extreme'' stars, in agreement with results from Milky Way GCs
(e.g., \citealt{Carretta2010}).  Future observations of more clusters
and more elements will enable trends to be investigated as a function
of other cluster parameters.

\section{The Power of IL Spectroscopy}
IL spectroscopy of GCs has provided valuable constraints on the
assembly of M31's halo, particularly its accretion history.  These
observations have yielded complementary information to resolved
photometry and spectroscopy of M31's brightest individual stars.
Future observations will expand the sample of clusters with known
metallicities, ages, and detailed chemical abundances, so that the
properties of their birth environments can be investigated more
thoroughly.  For observations beyond the Local Group, it is currently
very difficult to observe individual stars, and IL spectroscopy is
therefore the best option for probing the kinematics, chemical
abundances, and ages of distant systems (e.g., \citealt{Colucci2013}).
The advent of extremely large telescopes will allow individual stars
in distant systems to be observed and IL spectroscopy to push to
fainter targets. New developments in large spectroscopic surveys
(e.g., the Maunakea Spectroscopic Explorer; \citealt{MSE2019}) will
enable large numbers of GCs to be studied simultaneously, providing
more holistic characterizations of entire GC systems.  Ultimately,
advances in technology will enable distant galaxies to be studied at
the same level of detail that M31 can be studied today.

\end{document}